\documentclass[%
 reprint,
superscriptaddress,
 amsmath,amssymb,
 aps,
 floatfix,
 doublecolumn
]{revtex4-2}

\usepackage{graphicx}
\usepackage{bm}

\begin{document}
\title{Coexistence of trapped and flow-transported nuclei enables fast pigeon-post communication across multi-nucleated cell}

\author{Johnny Tong}
\thanks{These authors contributed equally to this work.}
\affiliation{%
 Technical University of Munich, TUM School of Natural Sciences, Department of Bioscience, Center for Protein Assemblies (CPA), Garching, Germany
}
\author{Kaspar Wachinger}
\thanks{These authors contributed equally to this work.}
\affiliation{%
 Technical University of Munich, TUM School of Natural Sciences, Department of Bioscience, Center for Protein Assemblies (CPA), Garching, Germany
}
\author{Fabian K. Henn}
\affiliation{%
    Technical University of Munich, TUM School of Natural Sciences, Department of Bioscience, Center for Protein Assemblies (CPA), Garching, Germany
}
\author{Nico Schramma}
\affiliation{%
    Institute of Physics, University of Amsterdam; Science Park 904, Amsterdam, The Netherlands
}
\author{Siyu Chen}
\affiliation{%
 Technical University of Munich,  TUM School of Natural Sciences, Department of Bioscience, Center for Protein Assemblies (CPA), Garching, Germany
}%
\affiliation{%
    Max Planck Institute for Dynamics and Self-Organization, 37077 G\"ottingen, Germany
}

\author{Karen Alim}%
 \email{k.alim@tum.de}
\affiliation{%
 Technical University of Munich,  TUM School of Natural Sciences, Department of Bioscience, Center for Protein Assemblies (CPA), Garching, Germany
}%
\affiliation{Max Planck Institute for Dynamics and Self-Organization, 37077 G\"ottingen, Germany}

\date{\today}
\begin{abstract}
\noindent Multi-nucleated cells exist in all domains of life, ranging from animals, plants and fungi to single-celled organisms such as the slime mold \textit{Physarum polycephalum}. The large cell size, in the case of \textit{Physarum} reaching centimeters and more, challenges the coordination of nuclei activity as signals need to cross large distances. In search for a mechanism for fast long-ranged communication among nuclei, we quantify nuclei dynamics and cytoplasmic flows in \textit{Physarum}'s tubular network. We observe nuclei in two interchangeable, dynamic states: mobile, flowing within the cytoplasmic shuttle flow, or trapped in the tube's porous cell cortex. As we find nuclei to accumulate at the tube's inner fluid-porous interface we theoretically explore and confirm, with physiological parameters, that slowing down of mobile nuclei during flow is sufficient for diffusible signal exchange between mobile and trapped nuclei. We analytically derive that communication akin to pigeon-post with mobile nuclei serving as pigeons shuttling between trapped nuclei acting as waypoints, gives rise to signaling velocities that account for the rapid intracellular reorganization observed in \textit{Physarum}. Since signal transfer by flow-transported nuclei outcompetes the mere diffusion of signals encoded in cytosolic proteins, pigeon-post communication surpasses alternative signaling mechanisms, even diffusive relay signaling up to twenty-fold in velocity. The key ingredients of pigeon-post communication, namely alternating flows and waypoints, exist in other multi-nucleated cells and may also be generalized beyond intracellular signaling.
\end {abstract}


\maketitle
\thispagestyle{empty}
Multinucleated cells are found in diverse forms of life, either within multicellular organisms as muscle \cite{bursztajn1989differential}, placental cells \cite{fogarty2011quantitative}, and cancer cells \cite{winkler2018harmful}, filamentous fungi \cite{roper2011nuclear, Mela2020}, or as single-celled organisms ranging from giant algae like \textit{Caulerpa} \cite{Arimoto2019, ranjan2015intracellular}, the malaria-causing \textit{Plasmodium falsiparum} \cite{bannister2000ultrastructure}, to slime molds like \textit{Physarum polycephalum} \cite{Sauer1982_book}.
Across such syncytial cells, defying classification as single-cellular nor multicellular organism, nuclei may encode spatially differentiated functions via differential mRNA transcription \cite{Kasuga.2008, Gerber2022, Bekker.2011,ranjan2015intracellular}. To achieve coordinated nuclear states, nuclei exchange cytosolic proteins as signals to synchronize transcription \cite{bataille2017dynamics} or initiate mitosis \cite{Chang.2013,Hayden.2022}. In contrast to very general small signaling molecules such as calcium, communication by signaling proteins is specific, triggering a chosen signal response only. Proteins smaller than $40\,\text{kDa}$ have free passage out of a nucleus through the nuclear pore complex \cite{wente2010nuclear} or by nuclear transporters \cite{markina2011nuclear, etxebeste2013cytoplasmic}. Bathed in common cytosol diffusible signals such as cytosolic proteins, but also mRNA, can be exchanged between nuclei for communication \cite{kloc2024syncytia}. Yet, the sheer size of syncytial cells, ranging up to centimeters, poses a fundamental challenge for nuclei communication via signaling molecules across the cell.

The so far fastest signaling mechanism for cytosolic proteins among nuclei is diffusive relay, where proteins diffuse between nuclei and trigger their own emission/activation at each newly reached nucleus such that the signal spreads collectively with a constant velocity of $0.6-2\,\mu \text{m/s}$ determined by the square root of signal diffusivity to signal doubling time \cite{Chang.2013, Hayden.2022,dieterle2020dynamics,Gelens.2014}. With diffusive relay, a signal can cross a distance of $10\,\text{mm}$ in about $1.5$ hours, beating pure diffusion by orders of magnitude, which would require more than $1400$ hours at a typical protein diffusivity of $10\,\mu \text{m}^2\text{/s}$. And yet $1.5$ hours is too slow to account for the rapid reorientation in intracellular concentration gradients in $10\,\text{mm}$ sized \textit{Physarum} cells, reportedly requiring only $10-30$ minutes \cite{Natsume.1993, Mori1986}. Although intracellular gradients reorientation in \textit{Physarum} is so far only determined by small signaling molecules, the stability of gradients before and after reorientation \cite{Natsume.1993} suggest signaling molecules to be downstream of specific transcriptional signals \cite{Milanese2019, Soboloff2012} that reoriented $3-9$ times faster than diffusive relay could account for.  

In \textit{Physarum}, cytoplasmic shuttle flows speed up intracellular transport   \cite{Kamiya1940}. Cytoplasmic flows arise because the cytoplasm-filled tubes making up its network-shaped body are lined with an actomyosin cortex \cite{Wohlfarth-Bottermann1979, Radszuweit2011,Oettmeier2017,Boussard.2021,Rieu.2015}, which rhythmically contracts \cite{Julien2018} driving network-spanning shuttle flows \cite{Alim2013}. Due to the flows, radial diffusion across longitudinal streamlines gives rise to the Taylor dispersion effect \cite{Taylor.1953,Aris.1956} enhancing effective longitudinal molecular diffusivity \cite{Marbach:2016}. Yet, for slowly diffusing proteins, the Taylor dispersion effect only unfolds over longitudinal scales larger than the  $10\,\text{mm}$ sized \textit{Physarum} cells considered here. Therefore, Taylor dispersion cannot speed up diffusive relay. Intuitively, one would expect that the transport of nuclei within the shuttle flow brings distant nuclei together. Yet, the time-reversibility of the low Reynolds number shuttle flows \cite{Alim2013,Kamiya1940} reverts nuclei back to their original position after each shuttle flow period of about $120$ seconds \cite{Haupt.2020}. Thus, flow-transported nuclei, but also diffusible signals, like proteins, emitted by nuclei are entrained with the flows and never leave their fluid neighborhood \cite{Purcell.1977}. However, nuclei can also get trapped within the porous gel cortex lining the tube wall \cite{Gerber2022}, such that time-reversibility is broken between trapped and flow-transported nuclei. Now flow-transported nuclei can, in fact, approach initially distant trapped nuclei. 

The two dynamic states of nuclei, mobile within flows and trapped, suggest an analogy to historic long-distance communication via pigeons in between waypoints. In the historic communication scheme, the pigeons' characteristic to return home was used to individually train pigeons to fly between two homes chosen as designated communication waypoints \cite{NAP1831}. Attaching messages to the mobile pigeons and handing over, i.e.~relaying, messages in between pigeons at waypoints, messages were bi-directionally spread over networks of waypoints \cite{Dash2022}. Although the historic pigeon-post communication was invented by humans, it does not rule out that nature also uses this fast, specific, and bi-directional communication scheme relying on mobile signal carriers, say advected by fluid flows, and waypoints, say trapped receivers and emitters of messages.
\begin{figure*}[!t]
  \centering
  \includegraphics[width=17.8cm]{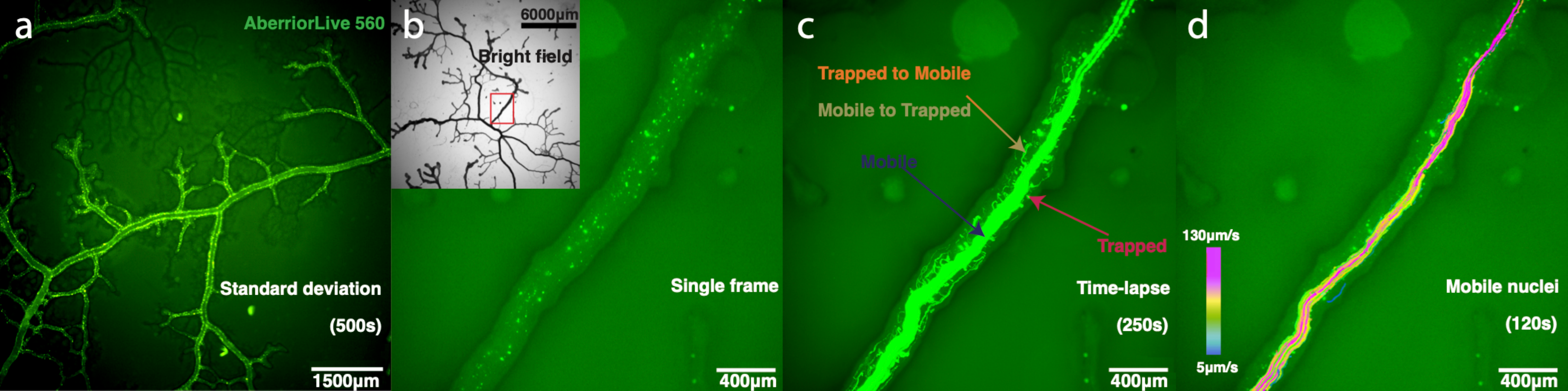}
  \caption{\textbf{Visualization of individual nuclei dynamics in \textit{Physarum}'s tubes reveals nuclei in two dynamic states: trapped and mobile} (a) Network-wide nuclei population imaged in $500s$ long time-lapse of a \textit{Physarum} network with nuclei labeled by double-stranded DNA specific AbberiorLive560 highlighting nuclei dynamics with the standard deviation of pixel intensity over 1000 frames. (b) Single frame close-up of labeled nuclei in a single tube within an entire network (inset). (c) Summation over $250s$ long time-lapse images of labeled nuclei in (b) reveals trapped and mobile dynamic states of nuclei as well as transitions between these states. (d) Trajectories of tracked mobile nuclei in one shuttle flow period of 120s color-coded by their maximal velocities. 
  }
  \label{fig:1}
\end{figure*}

We experimentally follow and quantify nuclei and cytoplasmic flows within the tubular network of \textit{Physarum} in search for a fast long-range communication mechanism accounting for its rapid intracellular re-organization \cite{Natsume.1993, Mori1986}. We find nuclei in two states: 1) mobile, flowing with the cytoplasmic flow, and 2) trapped, immobilized in the porous tube walls, as well as transitions between both states. Mobile nuclei only travel a fraction of the overall network size as they are entrained in the back-and-forth shuttle flow, which we quantitatively describe as Poiseuille flow with slip velocity at the tube's inner boundary. Notably, we observe that both mobile and trapped nuclei localize near the interface of cytoplasm and porous tube wall, bringing them in close proximity. We analytically derive that \textit{Physarum}'s physiological parameters allow for the exchange of specific diffusible signals encoded in cytosolic proteins between mobile and trapped nuclei due to the proximity of mobile and trapped nuclei and the slip flow at the inner tube boundary. Local diffusible signal exchange enables fast, long-ranged communication as mobile nuclei pigeon-like reverse their motion periodically with the shuttle flow and can thereby relay signals in between trapped waypoint-like nuclei during their extended proximity at flow reversals. Collective signal velocities from this flow-enhanced relay outcompete diffusive relay by a factor of $7$ to $28$. We provide the physical foundation of fast, specific, and bi-directional signaling with ``pigeon-post'' advection-diffusion relay rooted in the dual dynamic state of senders, trapped and advected by fluid flows,  which we envision to be of general applicability in syncytial cells and beyond.
\section*{Results}
\subsection*{Two co-existing nuclei states interact with cytoplasmic flow}
\begin{figure*}[t]
  \centering
  \includegraphics[width=17.8cm]{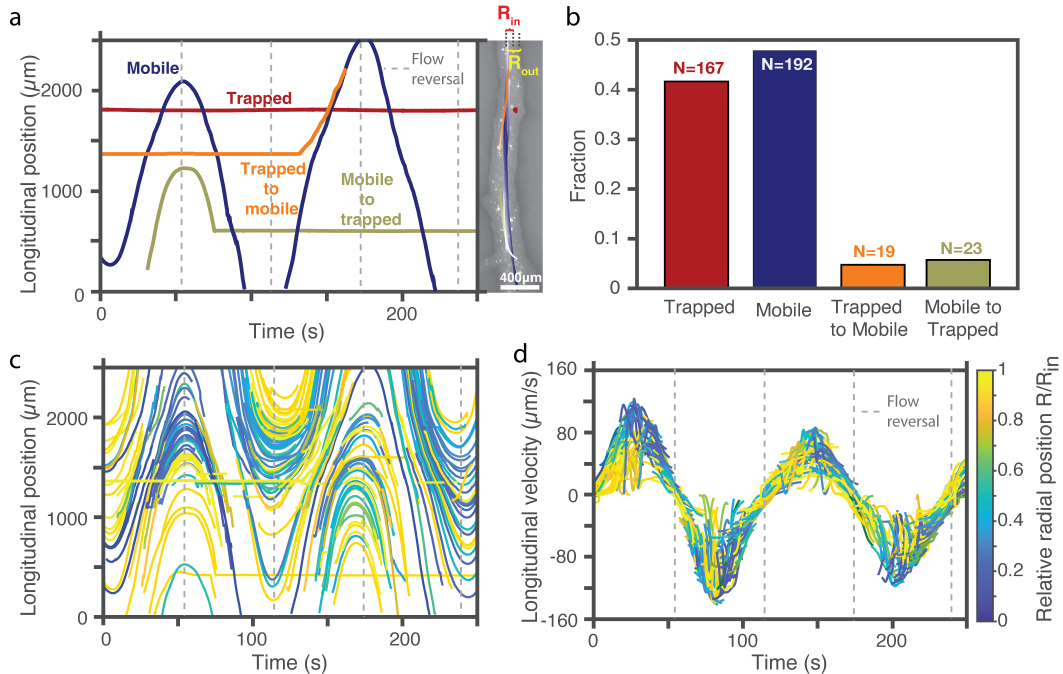}
  \caption{\textbf{Two dynamic states of nuclei are interchangeable: trapped - within the porous cortex lining the tube wall on the inside or mobile - advected within the cytoplasmic flow} (a) Trajectories exemplifying two nuclei states, trapped and mobile, and dynamic transitions between the two states. Right: Overlay of trajectories on fluorescence image. Here the exemplary track of the mobile nucleus is inferred by matching the nuclei in size and intensity during re-entry. (b) Statistics of states and transitions among a total count of 359 nuclei tracked over one shuttle flow period of 120s. N refers to the instances of tracked nuclei. (c) Longitudinal displacements of mobile nuclei follow back-and-forth shuttle flow. (d) Longitudinal velocities of nuclei, color-coded by their radial distance to the tube’s center line, oscillate sinusoidally reminiscent of Poiseuille-shaped peristaltic flows. The closer to blue in color, the closer the nucleus is to the center line. Data Set 2.}
  \label{fig:2}
\end{figure*}

To visualize trajectories of individual nuclei within the tubes of \textit{Physarum}, we fluorescently label nuclei by microinjecting Abberior LIVE $560$ DNA dye (Fig.~\ref{fig:1}, SI Movie 1-5, details see \textit{Materials and Methods}). Nuclei staining reveals a large number of distributed nuclei that follow long trajectories if centered in the tube, with shorter and shorter trajectories the more off-centered they are (Fig.~\ref{fig:2}c). In fact, in a large fraction of the tube close to the tube's outer wall, nuclei almost do not move at all; they rather seem to be trapped in the porous cortex lining the tube wall on the inside and separating the tube into a fluid inner and a porous outer part. 

To follow in detail individual nuclei dynamics we project trajectories onto the tube's center-line. Thereby we map out a nucleus' longitudinal position along the tube and its radial position relative to the tube's center-line (\textit{Materials and Methods}). Focussing on the longitudinal dynamics along the tube, we identify two distinct states of nuclei: trapped nuclei immobilized by the gel-like porous cortex lining the tube wall and mobile nuclei advected by the shuttle flow within the tube's inner radius (Fig.~\ref{fig:2}a). Taking statistics over 359 nuclei within one shuttle flow period ($T=120\,\text{s}$), we find trapped and mobile nuclei states with roughly equal probability but also around $10\,\mathrm{\%}$ of nuclei to transition from trapped to mobile or from mobile to trapped state (Fig.~\ref{fig:2}b). Trapped nuclei can be released, and mobile nuclei can be trapped again during the tube's cross-sectional contraction and subsequent relaxation (Fig.~\ref{fig:2}a, SI Movie 6-8). The flow-driven interchangeability between nuclei's dynamic states suggests that both trapped and mobile nuclei are biochemically active independent of temporary spatial entrapment. 

Mobile nuclei move collectively back and forth within the oscillatory cytoplasmic flow (Fig.~\ref{fig:2}c, d). However, nuclei visit only a fraction of the network before they return to their initial position after every period due to the time-reversibility of the low Reynolds number flow. Here, most nuclei travel a maximal distance of $2.5\,\mathrm{mm}$ within a network of $20\,\mathrm{mm}$ size (Fig.~\ref{fig:2}c). Nuclei positioned radially in the center of the tube move fastest and thus farthest. In fact, nuclei's longitudinal velocities follow a Poiseuille flow profile along the radial direction (Fig.~\ref{fig:2}d,~\ref{fig:3}b). Next, we quantify the cytoplasmic flows to determine the specifics of the nuclei's advection.
\subsection*{Porous tube walls promote Poiseuille flow with slip boundary}
\begin{figure*}[!t]
  \centering
  \includegraphics[width=17.8cm]{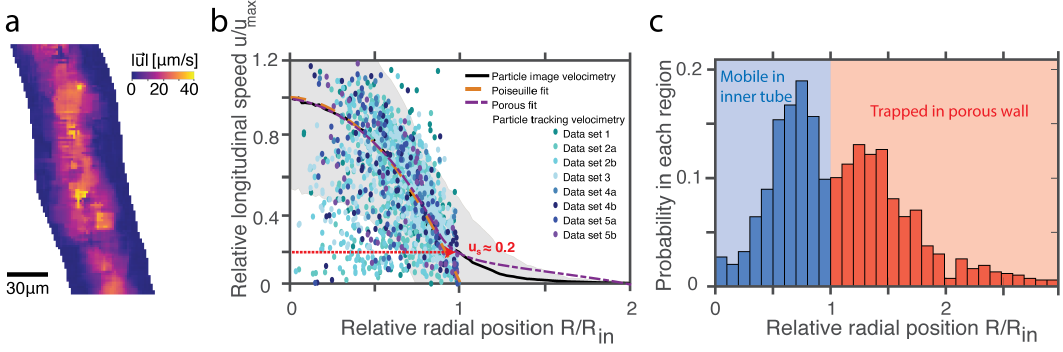}
  \caption{\textbf{Poiseuille-flow with slip boundary at inner tube's fluid-porous interface advects radially off-centered distribution of nuclei} (a) Flow profile along a tube from particle image velocimetry (PIV) at a time point of maximal flow velocity. Regions of zero flow velocity are represented in white. (b) Radial flow profile averaged along the tube shown in (a) together with individual nuclei velocities across data sets normalized by the maximal velocity of their sinusoidally varying velocity (See Fig.~\ref{fig:2}c). Flow profile with slip velocity amounting to $20\,\%$ of the tube's maximal velocity at the interface of fluid/inner to porous/outer tube at $R_{\text{in}}$ best captures PIV data. Grey area marks the standard deviation of PIV, representing variations in flow profile due to alterations in curvature and radius along the tube in (a). (c) Time-averaged probability distributions of mobile nuclei in the inner tube and trapped nuclei in the porous outer tube wall. Nuclei in the inner region are radially off-centered, peaked at $R=0.75R_{\text{in}}$. Nuclei in the porous outer tube wall concentrate close to the slip boundary.}
  \label{fig:3}
\end{figure*}

We use two independent methods to quantify fluid velocity and nuclei velocity. First, to quantify fluid velocity, we employ particle image velocimetry (PIV) with autofluorescent pigments (\textit{Materials and Methods}) (Fig.~\ref{fig:3}a). Second, to quantify nuclei velocity, we tracked the velocities of 1816 mobile nuclei across five independent recordings. Note that due to the large size of nuclei, we expect nuclei velocities to differ from PIV measurements, particularly in the vicinity of the porous cortex, where we observed that nuclei get trapped. 
This cortex is located at a radial distance $R_{\text{in}}$ from the tube's center. We measure this inner radius from the average maximal radial position of mobile nuclei trajectories to range between $30\,\%$ to $50\,\%$ of the total tube radius $R_{\text{out}}$  across all five data sets with $50\,\%$ the most common inner radius  (Fig.~\ref{fig:1}c).

Within the inner radius, nuclei velocities $u_{\text{n}}$ perfectly follow a Poiseuille flow profile (Fig.~\ref{fig:3}b). PIV data measuring the fluid itself, at first sight, also suggests a Poiseuille-like flow, in particular, due to the large standard deviation arising from variations in curvature and radius along the tube segment we average over. Yet, the fluid velocity does not decay to zero at the interface between the inner and outer tube in agreement with previous, independent measurements \cite{Haupt.2020}. That is, the no-slip boundary condition at the inner radius does not hold. Rather, the flow extends into the porous tube wall, albeit at decaying flow magnitude. Notably, the fluid velocity at the interface of inner and outer tube $R_{\text{in}}$ is $20\,\text{\%}$ of the maximal flow velocity in the center of the tube (Fig.~\ref{fig:3}a,b). 

We model the fluid flow profile across the inner and outer tube as a two-phase viscous and porous flow (Fig.~\ref{fig:3}b, \textit{SI Appendix Fig.~S4}) to gain a physical intuition on how a slip velocity may emerge due to the porous structure of the gel-like cortex of \textit{Physarum} \cite{Oettmeier2018,Guy2011}. In brief, as the lubrication approximation applies to the long and slender tubes considered here, the fluid phase is described by the Stokes equation, whereas the porous phase is modeled by the Brinkman equation. Both fluid and porous flow are solved consistently by applying continuous boundary conditions for flow velocity and strain rate at the inner tube fluid-porous interface $R = R_{\text{in}}$, finite flow velocity for all $R$, and a no-slip boundary condition at the outer tube wall $R/R_{\text{out}}=1$ (See \textit{SI Appendix, Text 4}) to arrive at a flow velocity of
\begin{align}
\begin{split}
u(r\leq\hat R) &= u_{\text{max}}\left(
1-\frac{r^2}{\hat{R}^2}\left(1-u_s\right)\right) \\ u(r\geq\hat R)&=\frac{4u_{\text{max}}(1-u_s)}{\hat{R}^2}\left(B_0 I_0(\alpha r) + B_1 K_0(\alpha r) + \frac{1}{\alpha^2}\right)
\end{split}
\label{eqn_porousflow}
\end{align}
 where $\hat R = R_{\text{in}}/R_{\text{out}}$ and $r=R/R_{\text{out}}$, $u_{\text{max}}=u(0)$ maximal velocity at the tube's center and $u_s = u(\hat R)/u_{\text{max}}$ the normalised slip velocity at the fluid-porous interface. $I_0$ and $K_0$ are zeroth order Bessel functions of the first and second kind. The parameter $\alpha$ is inversely related to the permeability $k=R_{\text{out}}^2/\alpha^2$ of the porous outer tube wall \cite{Guy2011,Singh2020} and $B_0=B_0(\alpha,\hat{R})$, $B_1=B_1(\alpha,\hat{R})$ are functions of $\alpha$ and $\hat R$ (See \textit{SI Appendix, Text 4}). The model fits our PIV data for a normalized slip velocity of $u_s = 0.2$ and permeability parameter of $\alpha = 27$ (Fig.~\ref{fig:3}b). Together with the outer tube radius $R_{\text{out}}\approx 40\,\mathrm{\mu m}$ for our PIV data set (Fig.~\ref{fig:3}a)  this results in a porous media permeability of $k \approx 2\,\mathrm{\mu m^2}$ (See \textit{SI Appendix, Text 4}). 
 This permeability is in line with large pores referred to as invaginations in \textit{Physarum}'s porous cortex \cite{ACHENBACH.1981} dominating permeability \cite{Nishiyama2017}.

Superimposing nuclei velocities on top of the PIV velocities as a function of relative radial position (Fig.~\ref{fig:3}b) further reveals a non-uniform distribution of nuclei with nuclei accumulation towards the inner fluid-porous interface. Mapping out the probability distribution of nuclei in the inner radius and in the porous outer wall separately (Fig.~\ref{fig:3}c) shows that the density of mobile nuclei peaks at around $0.75R/R_{\text{in}}$, relatively closer to the inner fluid-porous interface than to the tube's center. Time-lapse images substantiating denser nuclei flowing close to the inner fluid-porous interface (\textit{SI Appendix, Fig.~S2}) suggest the off-centered mobile nuclei positioning might arise due to margination of stiff, small nuclei by soft, large vacuoles in the cytoplasm \cite{Mueller.2014}. Note, that also trapped nuclei are concentrated close to the inner fluid-porous interface. Thus, the small spatial distance between mobile and trapped nuclei suggests that a diffusible signal could cross this short distance and facilitate communication between mobile and trapped nuclei. 
This communication would break the time-reversibility of the low Reynolds number flow, which otherwise constrains mobile nuclei and all diffusible signals they emit to remain in the same neighborhood. Although the shuttle flow deforms a mobile nucleus neighborhood as nuclei travel on different streamlines, time-reversibility ensures that after each shuttle period every nucleus returns to its original position. Only the two dynamics states, trapped and mobile, allow a nucleus to encounter a different neighborhood, namely the mobile or trapped nuclei it passes during the shuttle flow. 
\subsection*{Communication between trapped and mobile nuclei improved by slip boundary}
\begin{figure*}[!t]
  \centering
  \includegraphics[width=17.8cm]{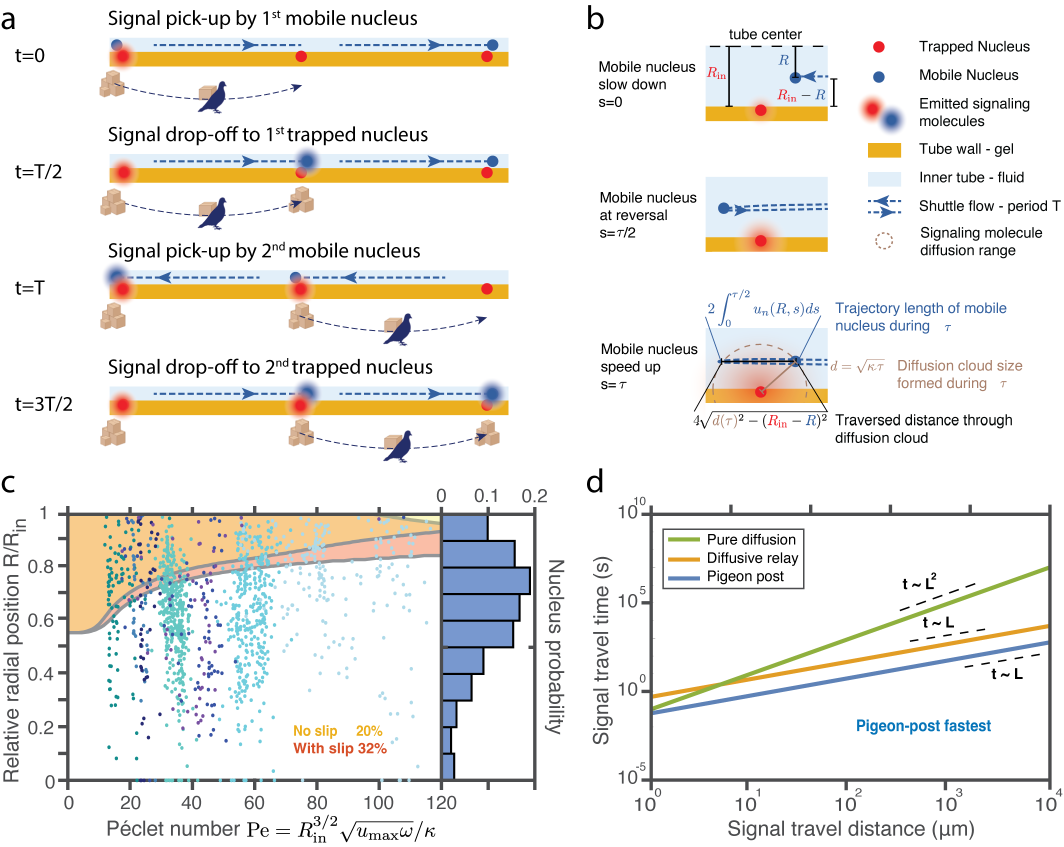}
  \caption{\textbf{Mechanism of pigeon-post communication between trapped and mobile nuclei enables fast long-range signaling} (a) Illustration of pigeon-post communication based on a subsequent successful communication via exchange of diffusible signaling molecules (cloud) between trapped (red) and mobile (blue) nuclei at flow reversals and the transport of the signal by mobile nuclei between trapped nuclei due to the shuttle flow. As mobile nuclei act as carriers (pigeons) along the tube and trapped nuclei act as waypoints (post office), signal exchange between mobile and trapped nuclei enables fast longitudinal relay communication over long distances. Note, for illustration we here sketch directed signal propagation only although the pigeon-post mechanism is intrinsically bi-directional. (b) Illustration of minimal condition for communication between a mobile nucleus and a trapped nucleus via emission of diffusible signaling molecules over time frame $\tau$ forming a diffusive cloud around one nucleus and reaching the other. Minimal conditions for communication are met if emission of signaling molecules starts $\tau/2$ before mobile nucleus' reversal in the shuttle flow. (c) Parameter range for fulfilled communication condition between mobile and trapped nucleus, without slip and with observed relative slip of $u_s = 0.2$, considering as time frame for the emission of signaling molecules $\tilde\tau\in[0.04,0.2]$. Tracked nuclei's physiological parameters $\{R, R_{\text{in}}, u_{\text{max}},\omega\}$ (blue dots corresponding to data sets as in Fig.~\ref{fig:3}b) suggest that one third of nuclei can participate in pigeon-post communication. Signaling molecule diffusivity set to $\kappa=10 \,\mu\text{m}^2\text{/s}$. (d) Comparison of scalings of signal travel time versus covered distance with pure diffusion and diffusive relay identifies pigeon-post communication as fastest. Molecule diffusivity of $\kappa=10 \,\mu\text{m}^2\text{/s}$, $u_{\text{diff.\, relay}}=2\,\mu\text{m}\text{/s}$, $u_{\text{pigeon-post}}=17\,\mu\text{m}\text{/s}$.}
  \label{fig:4}
\end{figure*}
If mobile and trapped nuclei were able to communicate via diffusible signals a relay signaling akin historic pigeon-post could unfold with mobile nuclei, serving as ``pigeons'', traveling large distances with the shuttle flow, passing on and receiving signals from trapped nuclei acting as ``waypoints''. Say a mobile nucleus just received a signal from a trapped nucleus (Fig.~\ref{fig:4}a, $t=0$). Ideally the mobile nucleus then travels the largest distance possible within the shuttle flow before dropping it off to another trapped nucleus, which would be when the mobile nucleus only picks up and drops off signals at its reversal points. Then the say rightward traveling mobile nucleus would drop off the signal at a trapped nucleus before the mobile nucleus reverses to travel leftward again (Fig.~\ref{fig:4}a, $t=T/2$). Having received the signal the trapped nucleus can multiply the signaling molecules while a formerly distant mobile nucleus to the right shuttles leftward and picks up the by now in sufficient amount emitted signaling molecules as it reverses in the trapped nucleus’ vicinity (Fig.~\ref{fig:4}a,  $t=T$) and heads off rightward again. Crucial for such pigeon-post communication to unfold is the exchange of a diffusible signaling molecule between a mobile and a trapped nucleus.

We next derive how mobile nucleus dynamics $u_{\text{n}}$, relative radial position of  mobile nucleus  $R/R_{\text{in}}$ and the time frame $\tau$ during which the sending nucleus emits sufficient signaling molecules constrain communication via exchange of diffusible signaling molecules. The oscillatory nature of the shuttle flow periodically slows down mobile nuclei before they briefly arrest, reverse, and slowly speed up in the opposite direction. We consider diffusible signal exchange around mobile nucleus reversal as it maximizes the signal travel distance in between ``waypoints'' and the time mobile and trapped nucleus spend in each others vicinity before the mobile nucleus is advected away again. The cloud of emitted signaling molecules deforms as they diffuse to streamlines of different velocities.  When the flow reverses, this deformation is reversed due to the time-reversibility of low Reynolds number flow. Thus, diffusion alone governs the shape of the signal cloud. Within time frame $\tau$, an emitted signaling molecule diffuses a distance $d(\tau)=\sqrt{\kappa \tau}$. We define a mobile and a trapped nucleus to be able to communicate by the exchange of diffusible signaling molecules if the mobile nucleus' trajectory allows them to be at a maximal distance $d$ from each other over the entire course of the time frame $\tau$. In other words, two nuclei need to spend enough time in each other's vicinity to develop diffusion clouds that overlap (Fig.~\ref{fig:4}b).  
For ease of argument, we here derive the communication condition from the perspective of a trapped nucleus emitting a diffusible signal to be received by a mobile nucleus, keeping in mind that the reverse role of emitter and receiver is included as well. The mobile nucleus' trajectory is slowest if we half our time frame  $\tau$ to be equally divided around the mobile nucleus' reversal. As the mobile nucleus slows down and speeds up on a streamline at radial position $R$, the time-averaged minimal spatial distance during time frame $\tau$ to a trapped nucleus at radial position $R_{\text{in}}$ is achieved if the trapped nucleus is positioned half-way along the mobile nucleus' longitudinal position before the reversal, which is revisited by the mobile nucleus again half-way after reversal, see Fig.~\ref{fig:4}b.  
The minimal conditions for communication are met if the mobile nucleus' trajectory at its radial distance of $R-R_{\text{in}}$ during time frame $\tau$ is shorter than length of the diffusion cloud. The longitudinal length of the diffusion cloud at the mobile nucleus radial position $R-R_{\text{in}}$ follows from solving the Pythagorean theorem with the diffusion cloud radius $d$ as the hypothenuse and $R-R_{\text{in}}$ as the short leg.
 Thus, the condition on the mobile nucleus' trajectory allowing for communication reads:
\begin{eqnarray}
  2\int_{0}^{\tau/2}u_{\text{n}}(R, s)ds\leq4\sqrt{d(\tau)^2-(R_{\text{in}}-R)^2}.
  \label{eqn_timeframe}
\end{eqnarray}
We describe the time-dependence of the nucleus' Poiseuille shuttle flow with sinusoidal oscillations of frequency $\omega$, i.e.,~$u_{\text{n}}(R,s)=u_{\text{max}}\left(1-R^2/R^2_{\text{in}}\right)\sin({\omega }t)$.
Note that not only nuclei but also diffusible signaling molecules are entrained in the fluid flow. With the finite slip velocity at the fluid-porous interface, trapped nuclei are localized, whereas the signaling molecules emitted by the trapped nucleus move with the slip velocity $u_{\text{max}}\, u_s\sin(\omega t)$ relative to the trapped nucleus. Therefore, the relative velocity $u_{\text{n,rel}}(R,t)$ of a mobile nucleus to the diffusing molecules emitted by a trapped nucleus at the inner tube boundary is reduced to $u_{\text{n,rel}}(R,t)=u_{\text{n}}(R,t) - u_{\text{n}}(0,t) u_s$. We incorporate the effect of the slip velocity for the signaling molecules by accounting for $u_{\text{n,rel}}(R,t)$ instead of $u_{\text{n}}(R,t)$ to solve the left-hand side of~\eqref{eqn_timeframe} (see \textit{SI Appendix, Text $5$}).  We find that the condition for communication \eqref{eqn_timeframe} between mobile and trapped nucleus depends on four non-dimensional parameters: the relative radial position of the mobile nucleus $R/R_{\text{in}}$, the time frame of signaling molecule emission rescaled by the intrinsic timescale for diffusion $\tilde\tau=\tau/t_{\text{diff}}$ with $t_{\text{diff}}=R_{\text{in}}^2/\kappa$,  the P\'eclet number $\text{Pe}=t_{\text{diff}}/t_{\text{adv}}$ with $t_{\text{adv}}=\sqrt{R_{\text{in}}/u_{\text{max}} \omega}$ the time scale for flow driven nuclei transport and $u_s$ the normalized slip velocity at the inner fluid-porous interface:
\begin{equation}
\tilde{\tau}-\left(1-\frac{R}{R_{\text{in}}}\right)^2\geq \left(\frac{\text{Pe}\cdot\tilde{\tau}}{4}\right)^4\left[1-\left(\frac{R}{R_{\text{in}}}\right)^2-u_s\right]^2.  
\label{communication_criteria}
\end{equation}

We map out the communication condition's \eqref{communication_criteria} phase space spanned by the mobile nucleus' relative radial position and P\'eclet number (Fig.~\ref{fig:4}c) while considering a signaling molecule emission time frame of $\tilde{\tau}\in[0.04,0.2]$, for different choices of  $\tilde{\tau}$ see Fig.~S5. The upper bound on the signal emission time frame is constrained by the time frame between signal pick up and drop off, i.e.~half the shuttle period of $T/2=60\,\mathrm{s}$, minus the minimal signal doubling time of $t_{\text{doubling}}=10\,\mathrm{s}$ \cite{Gelens.2014}. This yields $\tau=50\,\mathrm{s}$, corresponding to $\tilde{\tau}=0.2$ when normalized by $t_{\text{diff}}=250\,\mathrm{s}$. The lower bound follows from the maximal doubling time of $100\,\mathrm{s}$, which would  exceed the half period and give $\tilde{\tau}=0$. However, it has been shown that half a doubling time is already sufficient to initiate relay signaling \cite{dieterle2020dynamics}. We therefore set $\tau=10\,\mathrm{s}$, corresponding to $\tilde{\tau}=0.04$, as a lower bound. Note, that $\tau=10\,\mathrm{s}$ also corresponds to the time needed for a diffusion cloud to spread $0.2R/R_{\text{in}}$, the typical distance between trapped and mobile nuclei.  

At zero relative slip velocity, particularly nuclei either very close to the inner tube radius or at lower P\'eclet number, i.e.,~low flow velocity, have long enough time for communication while traversing the diffusion cloud.  However, taking into account the normalized slip velocity of $u_s=0.2$ as extracted from our data (Fig.~\ref{fig:3}b), communication is successful also at larger relative radial distance of at higher P\'eclet numbers up to $\text{Pe}=75$ (Fig.~\ref{fig:4}c). Due to the slip velocity, the diffusion cloud moves along with the mobile nuclei, thereby lowering the relative velocity of diffusing signaling molecules and mobile nuclei significantly and, thus, increasing the parameter space for successful communication. Intuitively, for an emitted signaling molecule cloud to become large enough to travel the distance between mobile and trapped nuclei, both nuclei must spend enough time in each other's vicinity. Slip velocity thus reduces the relative speed of the mobile nucleus and trapped nucleus' cloud to allow a longer time to transmit the diffusing molecules (Fig.~\ref{fig:4}a,c). In fact, one can locate experimentally tracked mobile nuclei within our theoretically predicted communication phase space by extracting all physiological parameters, i.e., individual nuclei's $R$ and $u_{\text{max}}$ as well as  $R_{\text{in}}$, $\omega$ and assuming a diffusivity of $\kappa = 10\,\mathrm{\mu m^2/s}$ for the signaling molecule matching proteins in size known from nuclei communication in mitotic waves \cite{wente2010nuclear} (SI Table 1).
As a result, we find that across 1816 tracked mobile nuclei from five independent datasets  $20\,\%$ of our tracked nuclei fulfill the communication condition without slip velocity, while with slip velocity, about $32\,\%$ of the mobile nuclei can engage in communication according to our calculations and measurements. The impact of the slip velocity is particularly significant at large P\'eclet number, where it doubles the amount of nuclei that meet our communication condition. Thus, both the radially off-centered distribution of nuclei peaked around $R=0.75 R_{\text{in}}$ (Fig.~\ref{fig:2}c) and the slip velocity at the fluid-porous inner tube interface (Fig.~\ref{fig:4}c) facilitate communication between mobile and trapped nuclei. Note that the large slip velocity allows for a higher maximal nuclei velocity and, thus, longer longitudinal trajectories of mobile nuclei per oscillation period, which implies that signals received by mobile nuclei can travel afar across the syncytial cell.
\subsection*{Pigeon-post mechanism supports fast long-distance signaling}
We next turn to estimate how quickly a signal could travel across the syncytial network by successive signal hand-over between mobile nuclei shuttling as ``pigeons'' between trapped nuclei acting as ``waypoints'' (Fig.~\ref{fig:4}c). Let's assume that the condition for communication between trapped and mobile nuclei, \eqref{communication_criteria}, is satisfied. Then a signal travels with a ``pigeon'' at $R/R_{\text{in}}$ between two reversals, i.e.,~within half a shuttle flow period $T/2$ the sinusoidally accelerating and slowing down nucleus travels a distance $\ell=u(R/R_{\text{in}})T/\pi$ before initiating over the course of the next half shuttle flow period $T/2$ the emission of the same signaling molecules at the ``waypoint'' (Fig.~\ref{fig:4}a). Thus, the ``pigeon-post'' travels with a velocity of $u_{\mathrm{pigeon-post}}=u(R/R_{\mathrm{in}})/\pi$. 
Repeated messaging between trapped nuclei acting as waypoints and mobile nuclei acting as pigeons along the syncytium speeds up the signal travel time with distance $L > \ell$ as 
\begin{equation}
t_{\text{pigeon-post}} = \frac{\pi L}{u(R/R_{\text{in}})}.
\label{pigeon_post}
\end{equation}
Note that in \textit{Physarum} nuclei are densely almost continuously distributed with a minimal inter-nuclei distance lower than $12\,\mathrm{\mu m}$ \cite{Gerber2022}. At such high density, the nuclei spacing does not constrain signal travel time. If a trapped nucleus is not at the ideal drop-off position a mobile nucleus must travel and emit slightly longer until reaching the next nucleus, at most $12\,\mathrm{\mu m}$ further along. Compared to the distance of $1.7$–$2\,\mathrm{mm}$ traveled during half a shuttle period, this mismatch reduces the effective signal travel distance by only about $1$–$2\,\%$. Further at this high nuclei density, there are about 50 mobile nuclei within every tube cross-section, implying that even if only $10\,\%$ of the mobile nuclei meet the communication condition, there are still 5 per cross-section. Also, trapped nuclei residence time exceeds the shuttle flow period (Fig.~\ref{fig:2}b), rendering signal travel time to be independent of the interchangeability of nuclei's dynamic state. 

In \textit{Physarum}, we observed mobile nuclei to accumulate around a radial position of roughly $R/R_{\text{in}}=0.75$ (Fig.~\ref{fig:3}c), just within the radial range where the communication condition between mobile and trapped nuclei \eqref{communication_criteria} is fulfilled (Fig.~\ref{fig:4}c). Thus, considering $u_{\text{n}}(R/R_{\text{in}}=0.75,t=T/4)=0.44\,u_{\text{max}}$ and $u_{\text{max}}=100-120\,\mathrm{\mu m/s}$ as observed experimentally (Fig.~\ref{fig:2}c,d) we find signal velocities in the range of $u_{\text{pigeon-post}}=14-17\,\mathrm{\mu m/s}$ only comparable in speed to signaling via unspecific tiny molecules as in calcium waves \cite{Stricker.1999}. Notably, pigeon-post's signal velocity is faster than diffusive relay signaling with cytosolic proteins, where signals only travel by diffusion between ``waypoints'', as the diffusive relay velocity is limited by the ratio of molecular diffusivity $\kappa$ and the emitted molecule doubling time $t_{\text{doubling}}$ with $u_{\text{diff.~relay}}=2\sqrt{\kappa/t_{\text{doubling}}}$ \cite{Gelens.2014}. Emitted molecule doubling time has been shown to increase at reduced emitter density and emission rate, as well as increased diffusion cloud volume and molecule sensing threshold \cite{dieterle2020dynamics}.
Taking as reference for the of signaling molecule doubling time  $t_{\text{doubling}}=10-100\,\text{s}$ as reported for excitable cells/nuclei \cite{Gelens.2014} and $\kappa =10\,\mathrm{\mu m^2/s}$ as before, diffusive relay velocities would only amount to $u_{\text{diff.~relay}}=0.6-2\,\mathrm{\mu m/s},$ $28$ to $7$-times slower than pigeon-post signaling, where the signal transport is by flow. The doubling time does not slow down pigeon-post signaling as the doubling time fits within the travel time of the mobile nucleus between two reversals where the signal is stored in the nucleus (mobile or trapped) but not needed to diffuse to another nucleus just yet. For our data on \textit{Physarum} the time between reversals amounts to  $T/2=60\,\text{s}$ (Fig.~\ref{fig:2}d). Furthermore, only a short distance between trapped and mobile nuclei and therefore small spatial volume needs to be covered by the emitted signaling molecule. Thus, the temporal alternation of signal communication at close distances and signal emission initiation in trapped nuclei or mobile nuclei while being transported seems particularly efficient.

As the signal travels with constant velocity, the signal travel time scales linearly in distance ($t\propto L$), therefore outcompeting diffusive messaging ($t\propto L^2$). Note that the enhancement of diffusivity by Taylor dispersion cannot speed up diffusivity for diffusive messaging or diffusive relay  here as it only acts in the physiological context of \textit{Physarum} for cytoplasmic proteins over a larger spatial distance than the $10^4\,\mathrm{\mu m}$ considered here.
Neither can pigeon-post be enhanced by  Taylor dispersion as mobile nuclei dynamics is governed by flow advection, crossing a distance of about $2\,\mathrm{mm}$ in half a shuttle flow period. Thus, for the physiological parameters of \textit{Physarum} we find  pigeon-post communication to outperform all other known competing signaling mechanisms. In particular, to cross a distance of $10\,\text{mm}$ pigeon-post would only requires $10$ minutes, just within the short time frame of observed stimuli response in \textit{Physarum} \cite{Natsume.1993, Mori1986}.
\section*{Discussion}
We here derived a fast collective relay signaling mechanism, akin to pigeon-post, based on signal activators existing in two dynamic states: 1) mobile nuclei, here advected within shuttle flows, as fast pigeon-like signal carriers and 2) trapped nuclei acting as waypoints for signals. As the resulting collective signal velocity is governed by fluid flow velocity, the pigeon-post mechanism is fastest for specific and bi-directional nuclei communication via cytosolic proteins. Even when considering the fast diffusivity of tiny molecules such as calcium and their enhancement by Taylor dispersion, the speed of pigeon-post communication is still comparable to such tiny molecule's diffusive relay yet without sacrificing the bi-directionally and specificity.
Routed in our observation of interchangeable dynamic states of nuclei in multi-nucleate \textit{Physarum}, which breaks low Reynolds number time-reversibility, we find that pigeon-post signaling with the here quantified physiological parameters encompasses at least one third of \textit{Physarum}'s nuclei and reaches the bi-directionally, specificity and short time scales observed in \textit{Physarum}'s intracellular reorganization.

Although the lack of versatile genetic manipulation in \textit{Physarum} \cite{Schaap.2015} prevents direct visualization of nuclei signaling here, mitotic waves, established as fast signaling systems in other syncytial cells \cite{Chang.2013,Hayden.2022}, would be a great first target.
Nuclei divisions in \textit{Physarum} have long been thought to be synchronous \cite{Howard.1932}, yet closer analysis revealed that synchrony breaks down when cytoplasmic flows are interrupted \cite{Wolf.1982} or in very large networks of $5\,\text{cm}$ in size \cite{Guttes.1961}. Both observations are in line with pigeon-post communication, where cytoplasmic flow is crucial for fast signaling and crossing distances of $5\,\text{cm}$ may still require up to one hour, which is significant compared to the roughly 8-hour mitotic cycle in \textit{Physarum} \cite{Sauer1982_book}. 
Spatial patterns of mRNA \cite{Gerber2022} in \textit{Physarum} suggest spatial nuclei identity. In fact, the density of trapped nuclei not only varies spatially within a \textit{Physarum} individual but actively changes upon stimulation with a food source (SI Appendix Fig.~S3). Thus, it would be interesting to investigate if spatial inhomogeneities in flow magnitude, net mass flow during migration or coordination of trapping and untrapping of nuclei may generate and protect spatial inhomogeneities in nuclei identity despite the cytoplasmic flows.

The key property for pigeon-post communication, namely the mobile and trapped nuclei, is, in fact, found in other syncytial organisms. In fungal networks, the perforated but to varying degrees remaining cell wall, termed septum, compartmentalizes fungal tubes but permits bi-directional transport of nuclei along tubes as well as trapping of nuclei at septa walls \cite{Lew.2005, Lew.2011}. Although the origin of nuclei transport there relies on molecular motors, the challenge of a low Reynolds number environment persists, suggesting that pigeon-post communication may well be found across extended systems, irrespective of nuclei acting as signal activators. Even in uni-directional flow/transport of a signal carrier, pigeon-post's inherent characteristic of storing signals at waypoints may, for example, permit long-distance communication when signal lifetime is limited. The mechanistic insight that two dynamic states of signal activators outcompete diffusive signal transport between signal activators may further hit fertile ground in the synthetic design of biological function in smart materials and autonomous soft robotics. 
\section*{Materials and Methods}
\subsection*{Cell culture}
\textit{Physarum} networks were cultured from microplasmodia developed from sclerotia (Carolina Biological Supply) following the protocols of Ref.~\cite{Bauerle:2017}. Specifically, the organism was cultured as microplasmodia in $100\,\mathrm{mL}$ of liquid culture, consisting of a $50\,\mathrm{\%}$ semi-defined medium (SDM) and $50\,\mathrm{\%}$ balanced salt solution (BSS), supplemented with hemin $5\,\mathrm{g/L}$, streptomycin, and penicillin \cite{daniel1961pure, daniel1962hematin}. Cultures were agitated at $120\,\mathrm{RPM}$ and maintained at a steady $25\,^{\circ}\mathrm{C}$, with the medium being replaced three times per week for optimal growth. Per the intended network, $100-200\,\mathrm{\mu L}$ of microplasmodia were let to settle in the pipette and placed in small droplets on a substrate plate with Phytagel\textsuperscript{TM} (Gellan Gum) $12\,\mathrm{g/L}$ to reduce autofluorescence. Microplasmodia fused and were kept at $25\,^{\circ}\mathrm{C}$ to grow into networks of $1-2\,\mathrm{cm}$ in size within $16-24$ hours.
\subsection*{Staining}
Microinjection was performed using an Eppendorf Femtotip\textsuperscript{TM} needle ($15\,\mathrm{{\mu}m}$ tip diameter) and a Microloader Pipette tip ($2-20\,\mathrm{{\mu}L}$ ). The dye, Abberior LIVE $560$ (Excitation max: $561\,\mathrm{nm}$, Emission max: $584\,\mathrm{nm}$), was dissolved in Dimethyl Sulfoxate (DMSO) and diluted in distilled water to a $10\,\mathrm{{\mu}M}$ concentration. Junctions in the plasmodial network were predominantly chosen as injection sites, allowing microinjection in the direction of the accelerating shuttle flow. Between three and six injections were microinjected for each dataset. Immediately after injection, the plate was placed in a $25\,^{\circ}\mathrm{C}$ incubator for about $40\,\mathrm{min}$ before imaging the specimen, preferably far from the injection site. For colocalization experiments to validate our staining method, $50\,\mathrm{{\mu}M}$ concentrated Abberior LIVE $560$ was microinjected and the same plasmodium was immersed in $20\,\mathrm{{\mu}M}$ concentrated Hoechst 33342 nuclear dye (Excitation max: $361\,\mathrm{nm}$, Emission max: $486\,\mathrm{nm}$).
\subsection*{Imaging}
For imaging at $12\,\mathrm{Hz}$, a Zeiss Axio Zoom V$16$ stereo microscope equipped with a Zeiss Plan Neofluar Z $1.0\,\mathrm{x}$ objective was used, along with a Hamamatsu Orca Flash $4.0$ V$2$ CMOS camera. Fluorescence imaging used filter cubes used for Abberior Live, Ex: $605/70\,\mathrm{nm}$, Em: $550/25\,\mathrm{nm}$ and for Hoechst  Ex: $365\,\mathrm{nm}$, Em: $445\,\mathrm{nm}$. For detailed colocalization analysis, see SI Appendix Text 1. Image acquisition was done using the ZEN Blue Edition software (Version $3.2$) by Zeiss.
Particle Image Velocimetry for auto-fluorescence particles was conducted in a confocal spinning disk microscope (Yokogawa CSU-X1), with EMCCD Camera (Andor iXon Ultra $897$) at a frame rate of $35\text{ms/frame}$ and a pixel size of $0.4\text{$\mu$ m}$ at $40\text{x}$ magnification using a Nikon $40x/0.6$ Luc Plan FL N, by exploiting auto-fluorescence of pigments in \textit{Physarum} (Em $488$/Ex $525$).
\subsection*{Data analysis}
Our data comprises five independent datasets, each $20\,\mathrm{min}$ long. To track the fluorescently labeled nuclei, first, the rolling ball function in FIJI is used for background subtraction. Nuclei positions and shapes are identified in each frame by spot detection and tracked across frames using a Linear Assignment Problem-based method \cite{jaqaman2008robust}. The spot detection algorithm is based on a Laplacian of Gaussian filter and a region formation algorithm using Marker-Controlled Watershed \cite{Aragaki2022}. To link spots between frames and prevent crosstalk of nuclei trajectories, we customized the cost function of linking in penalizing large angles between two consecutive links. We discard trajectories that last less than three frames.

The analysis of nuclei trajectories involves a combination of functions to separate nuclei motion in radial and longitudinal directions: First, we combine the two bright field recordings of each network before and after fluorescence recordings into a mask by Gaussian filtering and thresholding using a customized MATLAB code. Then the tube is skeletonized to obtain the tube center line. Subsequently nuclei positions and velocity vectors are projected to the center line via the spline-fitting exchange function distance2curve \cite{D'Errico} to obtain each nucleus radial and longitudinal position over time. Longitudinal nuclei velocities were calculated by spline-fitting of the longitudinal nuclei trajectories and subsequent numerical differentiation.

\begin{acknowledgments}
We thank Marcus Roper, Daniel E.~Rozen, Cathleen Broersma and Martin Lenz for fruitful discussions. We thank Christian Westendorf for microscopy support in the early stages of this work.
This work was supported by the Max Planck Society, the Human Frontier Science Program Organization through Research Grant number RGP0001/2021 and by the Federal Ministry of Education and Research (BMBF), the Free State of Bavaria under the Excellence Strategy of the Federal Government and the Länder for the TUM Innovation Network ``Robot Intelligence in the Synthesis of Life'' (K.A.) and the Deutsche Forschungsgemeinschaft DFG through INST 95/1634-1 FUGG.
\end{acknowledgments}


%

\end{document}


\title{\large Supplemental Materials: Coexistence of trapped and flow-transported nuclei enables fast pigeon-post communication across multi-nucleated cell}

\author{Johnny Tong}
\thanks{These authors contributed equally to this work.}
\affiliation{%
 Technical University of Munich, TUM School of Natural Sciences, Department of Bioscience, Center for Protein Assemblies (CPA), Garching, Germany
}
\author{Kaspar Wachinger}
\thanks{These authors contributed equally to this work.}
\affiliation{%
 Technical University of Munich, TUM School of Natural Sciences, Department of Bioscience, Center for Protein Assemblies (CPA), Garching, Germany
}
\author{Fabian K. Henn}
\affiliation{%
    Technical University of Munich, TUM School of Natural Sciences, Department of Bioscience, Center for Protein Assemblies (CPA), Garching, Germany
}
\author{Nico Schramma}
\affiliation{%
    Institute of Physics, University of Amsterdam; Science Park 904, Amsterdam, The Netherlands
}
\author{Siyu Chen}
\affiliation{%
 Technical University of Munich,  TUM School of Natural Sciences, Department of Bioscience, Center for Protein Assemblies (CPA), Garching, Germany
}%
\affiliation{%
    Max Planck Institute for Dynamics and Self-Organization, 37077 G\"ottingen, Germany
}

\author{Karen Alim}%
 \email{k.alim@tum.de}
\affiliation{%
 Technical University of Munich,  TUM School of Natural Sciences, Department of Bioscience, Center for Protein Assemblies (CPA), Garching, Germany
}%
\affiliation{Max Planck Institute for Dynamics and Self-Organization, 37077 G\"ottingen, Germany}

\maketitle

\section*{Supplementary Text 1: Colocalization test}
Co-localization was conducted using Abberior LIVE $560$ ($50\,\mathrm{{\mu}M}$) as the testing dye and Hoechst $33342$ (dissolved in DMSO and diluted in distilled water to a $20\,\mathrm{{\mu}M}$ concentration) as the control dye on plasmodial networks cultivated and grown following the protocol as described in Materials and Methods. Nuclei were labeled through microinjection of Abberior LIVE $560$. Exploiting cytoplasm containing nuclei oozing out of the injection site post-injection allowed better accessibility of nuclei for labeling by submerging in concentrated Hoechst $33342$. Furthermore, cross-fluorescence checks were performed, on both testing and control dye in two separate plasmodia, confirming Hoechst $33342$ labeled nuclei were not visible when imaging Abberior LIVE $560$ and Abberior LIVE $560$ labeled nuclei were not visible when imaging for Hoechst $33342$.

We analyzed the Pearson's correlation coefficients using the Coloc $2$ tool on FIJI and visualized the corresponding scatter plots. They showed a high correlation between Abberior LIVE $560$ and Hoechst $33342$ with overlap coefficient $R = 0.81$ and Pearson Coefficient $PCC= 0.70$.

\begin{figure}[h!]
	\centering
	\includegraphics[width=1\textwidth]{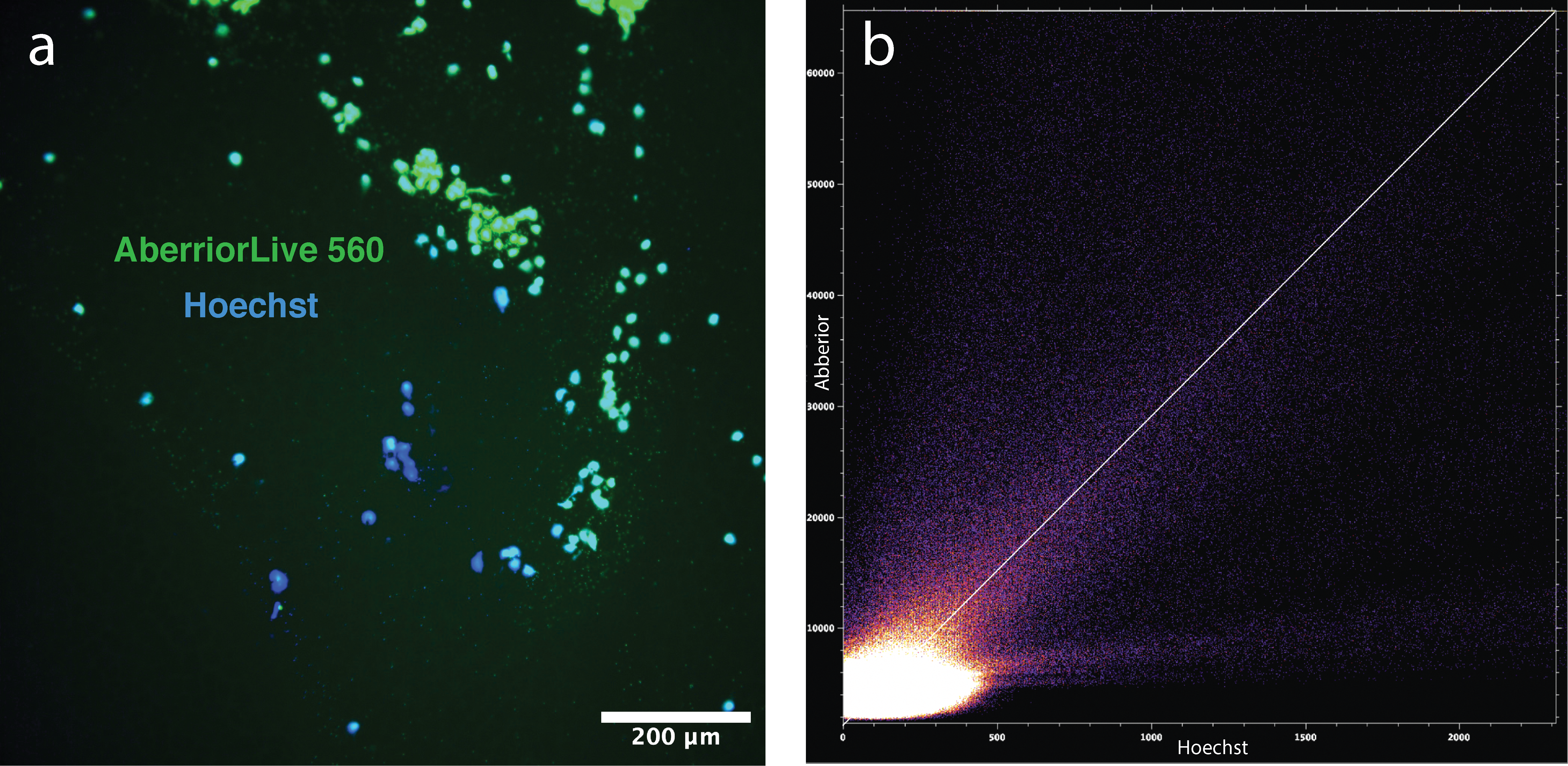}
	\caption{(a) Colocalization of co-labeled nuclei using microinjected Abberior LIVE $560$ (green) and submerging specimen in Hoechst $33342$ (Blue). (b) Linear regression fit shows high correlation of Abberior LIVE 560 and Hoechst with Pearson's R value of $0.7$}
	\label{fig:coloc}
\end{figure}

\newpage
\section*{Supplementary Text 2: Radial off-centered distribution}
\begin{figure}[h!]
	\centering
	\includegraphics[width=1\textwidth]{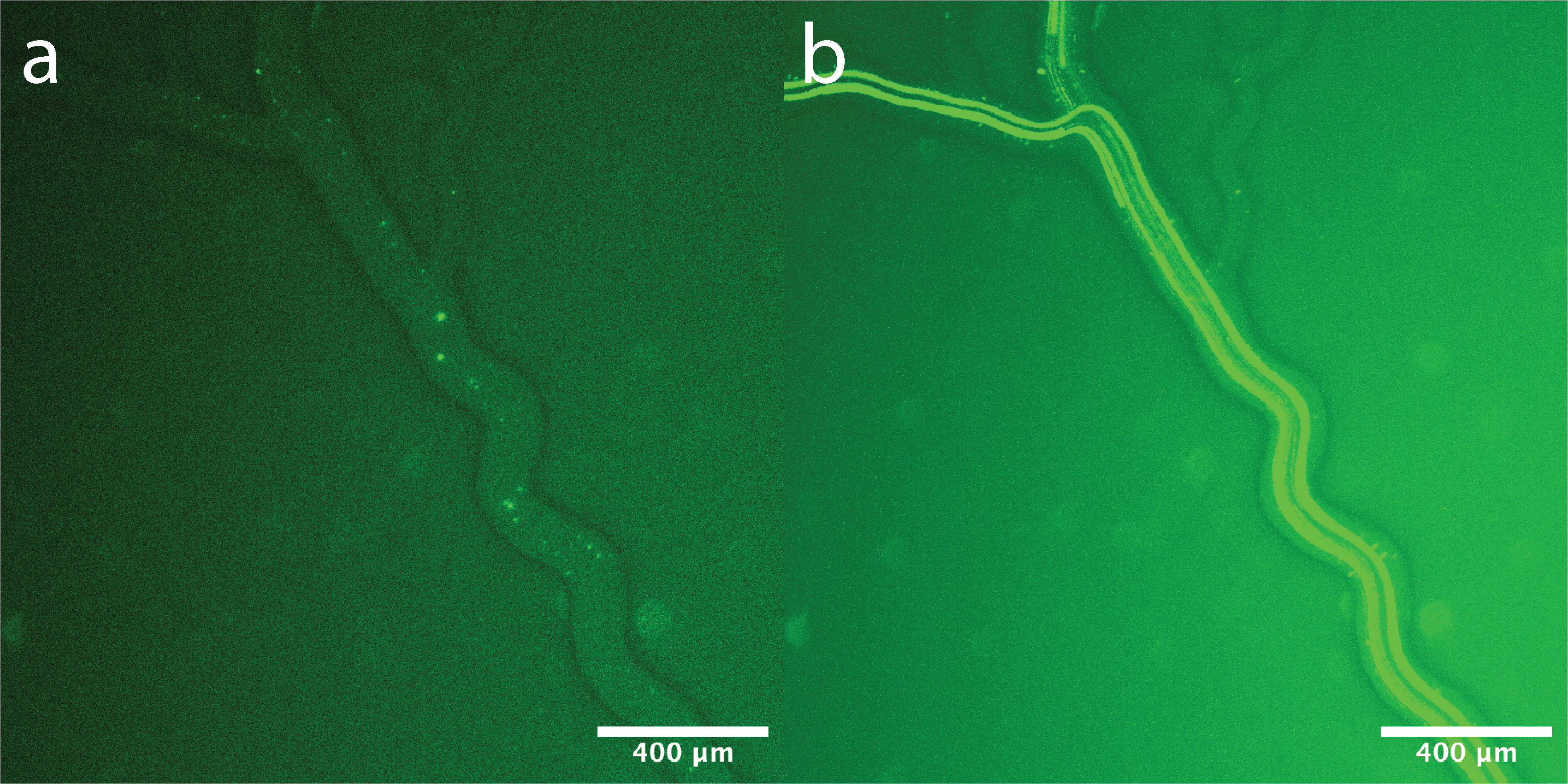}
	\caption{(a) Single frame individual labeled nuclei in a single tube with low density. (b) $250\,\mathrm{s}$ long time-lapse image of labeled nuclei in (a) shows that nuclei trajectories distribute radially off-centered.}
	\label{fig:radial}
\end{figure}

Off-centered distribution of nuclei accumulate close to the inner tube's fluid-porous interface is visualized with a representative image with lower nuclei density for higher contrast (Suppl.~Fig.~\ref{fig:radial}). Potentially the dense and stiff mobile nuclei get marginated by larger soft vacuole streaming within the cytoplasm \cite{Muller2014}. 

\newpage
\section*{Supplementary Text 3: Spatial variation of trapped nuclei affected by food stimulus}
\begin{figure}[h!]
	\centering
	\includegraphics[width=1\textwidth]{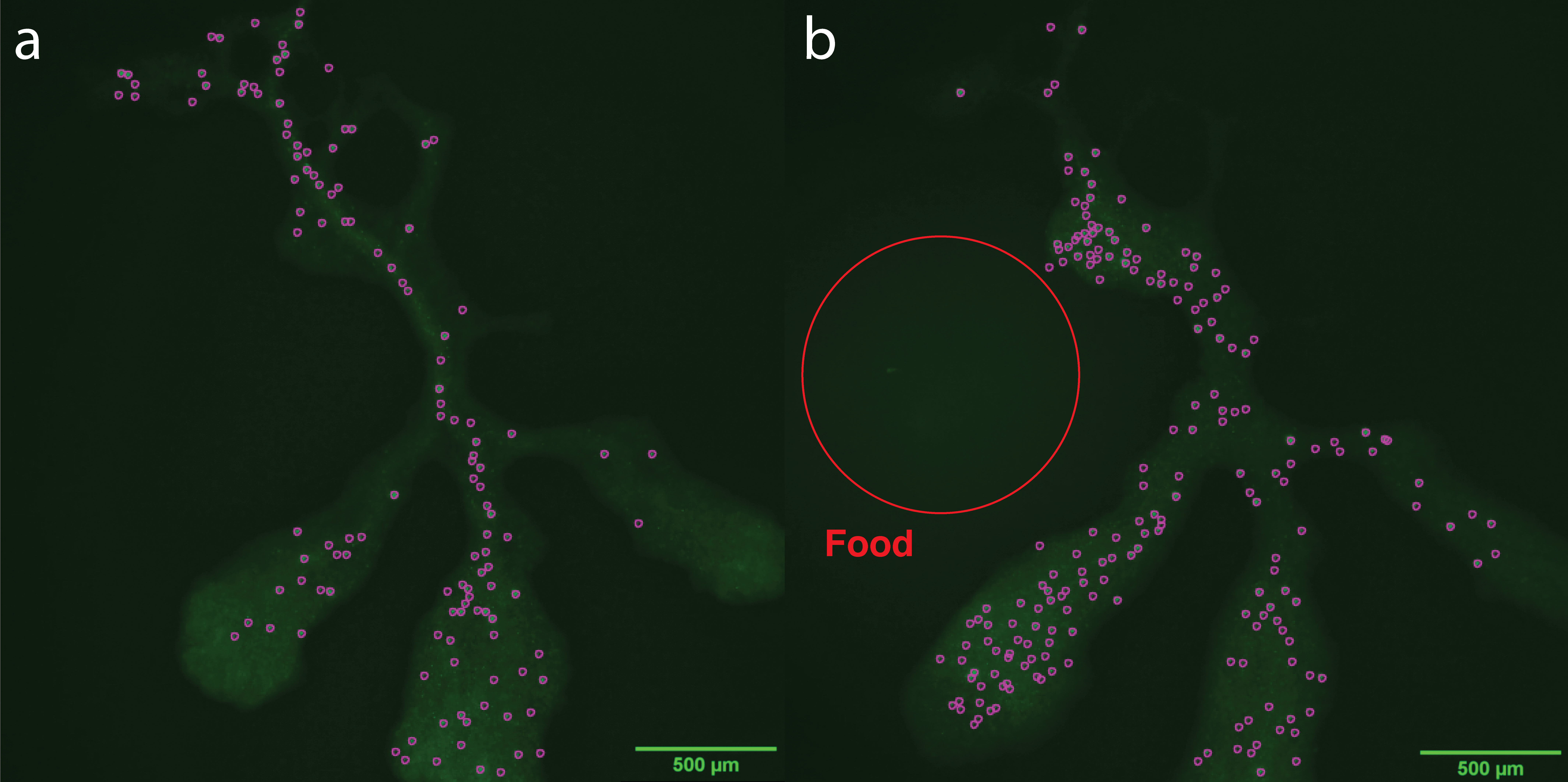}
	\caption{(a) Single frame with trapped nuclei highlighted by purple circles. (b) 1 minute after the food source (red circle) is added in the proximity of the \textit{Physarum} network the pattern of trapped nuclei changed.}
	\label{fig:food}
\end{figure}

Density of trapped nuclei may vary spatially within an individual \textit{Physarum} and is affected by food stimulus. To stimulate \textit{Physarum} by food gelous 0.8\% gellan gum is mixed with culture medium of 50\% BSS and 50\% SDM (see Materials and Methods). When locally stimulated with the food source by placing the food source in the proximity of the network, nuclei accumulate in the proximity of the food source within 1 minute (Suppl.~Fig.~\ref{fig:food}). The observation suggests that the density of trapped nuclei may be actively regulated by environmental stimuli.

\newpage
\section*{Supplementary Text 4: Flow in a tube with porous walls}
\label{sec:FlowPorousTube}
\begin{wrapfigure}{r}{0.5\textwidth}
	\centering
	\includegraphics[width=0.5\textwidth]{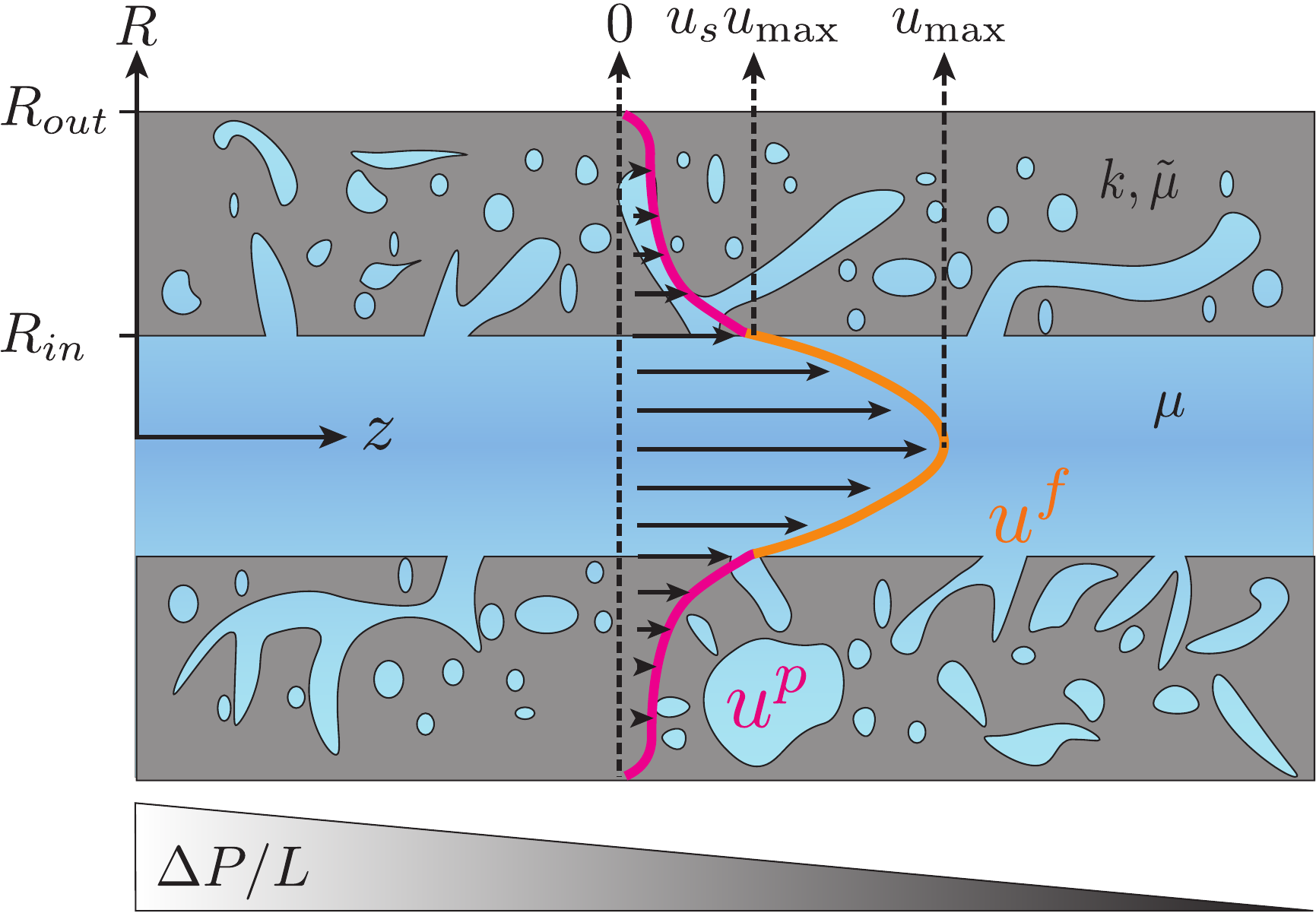}
	\caption{Schematic flow in a tube with Newtonian fluid of viscosity $\mu$ and effective viscosity and permeability $\hat\mu, k$ in the porous tube wall, for radii greater than $R_{\text{in}}$. The pressure driven tube flow is split into a porous tube flow $u^p$ and a free inner flow $u^f$, which matches in velocity and strain rate at the inner fluid-porous interface at at $R=R_{\text{in}}$.}
	\label{fig:schematicporousflow}
\end{wrapfigure}
We derive the flow field in the tubes of \textit{Physarum polycephalum} as a fluid enclosed by a porous shell. As the tube-wall was shown to be a porous medium \cite{Oettmeier2018}, we here consider a simple model for a pressure driven flow in a tube with a porous tube wall of porosity $k$ extending between the inner radius $R_{\text{in}}$ and the outer radius $R_{\text{out}}$ (Suppl.~Fig.~\ref{fig:schematicporousflow}). We assume, that the inner fluids' Newtonian viscosity and effective viscosity of the fluid within the porous wall are the same $\tilde\mu=\mu$ \cite{Brinkman1949} together with the tube diameter of $2R_{\text{out}}\approx 100\,\mathrm{\mu m}$ and the maximal speeds of $u_{\text{max}}\approx 1\,\mathrm{mm/s}$ we find a maximal Reynolds number of $\mathrm{Re}=2R_{\text{out}}u_{\text{max}}/\nu\approx0.1$ \cite{Alim2013} by assuming the kinematic viscosity of water $\nu$ which serves as a lower limit. Furthermore, we assume that the cytoplasm is incompressible and behaves as a purely Newtonian fluid in which the stress is proportional to the strain rate $\sigma=\mu \dot \gamma$. We work in a cylindrical coordinate system, with radial axis coordinate $R \geq0$, angular axis coordinate $\phi \in [0,2\pi)$ and tube axis $z\in \mathbb{R}$. Furthermore we assume a pressure drop only into the z direction along the tube axis: $\nabla p = \frac{\partial p}{\partial z}$ which actuates fluid motion only in z-direction $u_r=0 , u_\phi =0$. In \textit{Physarum} this pressure drop is induced locally by peristaltic pumping \cite{Julien2018}.
This leads to the Stokes equation for the cytoplasm in $0<r<R_{\text{in}}$ and to the Brinkman equation for the flow in the tube wall $R_{\text{in}} < R < R_{\text{out}}$ \cite{Neale1974}.
\begin{align}
\nabla \cdot \mathbf{u}^{f,p} = \frac{\partial u^{f,p}_z}{\partial z}  &= 0, \\
\mu \Delta \mathbf{u}^f = \mu \frac{1}{R}\frac{\partial}{\partial R} ( r\frac{\partial u^f_z}{\partial R} ) \hat{e}_z &= -\frac{\partial P}{\partial z} \hat{e}_z  ,\label{eq:stokes_r} \\
-\frac{\mu}{k} u^{p}_z + \mu \frac{1}{R}\frac{\partial}{\partial R} ( R\frac{ \partial u^{p}_z }{\partial R}) &= -\frac{\partial P}{\partial z}, 
 \end{align}
where superscript $f$ and $p$ indicate the inner flow or the flow in porous medium, respectively (Suppl.~Fig.~\ref{fig:schematicporousflow}).
For convenience, we will abbreviate $u:=u_z$. \\
Now the problem gets subdivided into three steps, first we find a general solution for the inner flow problem, then we show a solution for the Brinkman equation, and last but not least, we match both solutions by applying continuous boundary conditions at $R=R_{in}$. 
 
\paragraph{Solving the inner flow problem $u^f$}
The inner flow velocity follows the Stokes equation. 
As the pressure gradient is considered constant, we can simply integrate \eqref{eq:stokes_r} and find the typical Poiseuille-flow:
\begin{align}
u^f(R) = -\frac{1}{4} R^2 + A_0 \ln(R) + A_1,
\end{align}
with $A_0, A_1$ integration constants.
\paragraph{Solving the porous flow problem $u^p$}
As Singh et al.~\cite{Singh2020} already solved a similar, but more complicated variant of this problem we will follow their steps in this paragraph. By rescaling $R \to \frac{R}{R_{\text{out}}}$ and $u \to -\frac{\mu u}{R_{\text{out}}^2\frac{\partial P}{\partial z}}$ and by introducing $\alpha^2 =  R_{\text{out}}^2/k$ which compares the permeability $k$ to the tubes cross-sectional area we find:
\begin{align}
\dv[2]{u^p}{R} + \frac{1}{R} \dv{u^p}{R} - \alpha^2 u^p = -1,
\end{align}
which is the modified Bessel-equation with non-zero right-hand-side.
A linear combination of modified Bessel functions of zeroth order and first and second kind $I_0$ and $K_0$ solves this equation:
\begin{align}
u^p(R) = B_0 I_0(\alpha R) + B_1 K_0(\alpha R) + \frac{1}{\alpha^2},
\end{align}
with $B_0$ and $B_1$ constants. 
\paragraph{Boundary conditions}
At the interface between both media $\hat{R} = R_{\text{in}}/R_{\text{out}}$, we require that the flow speed and the strain rate $\frac{\partial u}{\partial_R}$ are continuous. The flow speed is finite for all $R$ and we impose a no slip boundary condition at the outer tube-wall $R=1$:
\begin{equation}
\frac{\partial u^f}{\partial R}(0)  = 0,\;\;\;\;u^f(\hat{R}) = u^p(\hat{R}),\;\;\;\; \frac{\partial u^f}{\partial R}(\hat{R}) = \frac{\partial u^p}{\partial R}(\hat{R}),\;\;\;\; u^p(1) = 0. 
\end{equation}
From the first boundary condition we find, that $A_0=0$, as the flow must be finite at $R=0$.
Next, we derive three equations for the three unknowns $A_1, B_0, B_1$:
\begin{equation}
\begin{split}
&B_0I_0(\alpha \hat{R}) + B_1 K_0(\alpha \hat{R}) + \frac{1}{\alpha^2}  +\frac{1}{4} \hat{R}^2 =  A_1,  \\
 &B_0\alpha I_1(\alpha \hat{R}) - B_1\alpha K_1 (\alpha \hat{R})= -\frac{1}{2} \hat{R} \quad \Leftrightarrow \quad B_0 = \frac{B_1K_1(\alpha \hat R)-\frac{\hat R}{2\alpha} }{I_1(\alpha \hat R)},\\
&B_0 I_0(\alpha) + B_1 K_0(\alpha) + \frac{1}{\alpha^2} = 0 \quad \Leftrightarrow \quad B_0 = -\frac{1+\alpha^2B_1 K_0(\alpha)}{\alpha^2 I_0(\alpha)},
\end{split}
\end{equation}
By combining the last two equations we find: 
\begin{equation}
\centering
\begin{split}
-[1&+\alpha^2 B_1 K_0(\alpha)]I_1(\alpha \hat R) = \left[\alpha^2 B_1K_1(\alpha \hat R) - \frac{\hat R \alpha}{2}\right] I_0(\alpha)\quad \\
 \Leftrightarrow  B_1 &= \frac{\hat R \alpha I_0(\alpha) - 2 I_1(\alpha \hat R) }{2\alpha^2 [K_0(\alpha) I_1(\alpha\hat R) +I_0(\alpha) K_1(\alpha\hat R)]} \\ 
 \Rightarrow\quad  B_0 &= \frac{K_0(\alpha)\left[2\frac{I_1(\alpha \hat R)}{I_0(\alpha)}-\hat{R} \alpha\right]}{2\alpha^2 [K_0(\alpha) I_1(\alpha\hat R) +I_0(\alpha) K_1(\alpha\hat R)]} - \frac{1}{\alpha^2 I_0(\alpha)}\\
 \Rightarrow \quad A_1 &= B_1\left[K_0(\alpha \hat R) - \frac{I_0(\alpha \hat R)K_0(\alpha)}{ I_0(\alpha)}\right] +\frac{1}{\alpha^2} - \frac{I_0(\alpha \hat R)}{\alpha^2 I_0(\alpha)}+ \frac{\hat R^2}{4}.
\end{split}
\end{equation}
This enables us to write down our final solution by re-substitution the rescaled variables:
\begin{align}
u(R) = \frac{R_{\text{out}}^2 \partial_z P}{4\mu} \begin{cases}
4A_1-\frac{R^2}{R_{\text{out}}^2} &  ,0 \leq R \leq R_{\text{in}} \\
4\left(B_0 I_0\left(\frac{\alpha R}{R_{\text{out}}}\right) + B_1 K_0\left(\frac{\alpha R}{R_{\text{out}}}\right) + \frac{1}{\alpha^2}\right) & ,R_{\text{in}} \leq R \leq R_{\text{out}}
\end{cases}
\end{align}

This flow accounts for the boundary layer between Poiseuille-like inner fluid flow, and the coarse-grained flow in the porous medium, which eventually decays to zero at the outer wall (see Suppl.~Fig.~\ref{fig:schematicporousflow}). The flow profile smoothly changes curvature and accounts for the apparent slip-like layer we observed using particle image velocimetry. The fitting parameter $\alpha = \frac{R_{\text{out}}}{\sqrt{k}}$ gives us an estimate for the tube permeability $k$. Notably, the pressure-driven velocity $U = \frac{R_{\text{out}}^2 \partial_z P}{4\mu}$ is not the same as the velocity at the center of the tube $u(0)$ due to the deviation of the flow profile from Poiseuille flow. We find: 
\begin{align}
u(0) & = 4U A_1 := u_{\text{max}}\\
u(R_{\text{in}}) &= U\left(4A_1-\hat{R}^2\right) := u_{\text{max}} u_s,
\end{align}
with the non-dimensional slip velocity \begin{align}
    \label{eq:uslip}
    u_s = 1 - \frac{\hat R ^2}{4A_1({\hat R}^2, \alpha)} \Leftrightarrow 4A_1 = \frac{\hat{R}^2}{1-u_s},
\end{align}
such that with $r=R/R_{\text{out}}$
\begin{align}
u(r) = u_{\text{max}} \begin{cases}
1-\frac{r^2}{\hat{R}^2}\left(1-u_s\right) &  0 \leq r \leq \hat R \\
\frac{4(1-u_s)}{\hat{R}^2}\left(B_0 I_0(\alpha r) + B_1 K_0(\alpha r) + \frac{1}{\alpha^2}\right) & \hat R \leq r \leq 1
\end{cases}
\label{eqn_porousflow}
\end{align}

\paragraph{Fit parameters}
Our fit parameters for the data shown in Fig.~3b of the main text are:
$u_{\text{max}}= 19.24\pm0.17\,\mathrm{\mu m/s}$, $\alpha= 27\pm2$, $\hat R= R_{\text{in}}/R_{\text{out}}=0.388\pm0.005$.
We find a permeability in the porous annulus $k=R_{out}^2/\alpha^2\approx 2\,\mathrm{\mu m}$.
From those we can calculate the slip length according to equation \eqref{eq:uslip}: $u_s\approx0.2$. 

\section*{Supplementary Text 5: Derivation of communication condition}

We formulate the conditions for communication between a mobile and a trapped nucleus by exchange of diffusible signals as follows. 
We assume that molecules from trapped nuclei diffuse a distance
\begin{equation}
    d = \sqrt {\kappa \tau},
\end{equation}
forming a diffusion cloud of radius $d$ in time $\tau$. 
The trajectory of a mobile nucleus at position $R$ traversing  at a radial distance $R_{\text{in}}-R $ away from the tube wall at radius $R_{\text{in}}$ for communication needs to be shorter than the distance it traces out in the diffusion cloud 
\begin{equation}
    4\sqrt{d(\tau)^2-(R_{\text{in}}-R)^2}=4R_{\text{in}}\sqrt{\frac{\kappa}{R_{\text{in}}^2} \tau-\left(\frac{R_{\text{in}}-R}{R_{\text{in}}}\right)^2}\geq 2\int_{0}^{\tau/2}u_{\text{n}}(R, s)ds,
\end{equation}
while traveling at velocity $u_{\text{n}}(R, t)$, assuming the mobile nucleus reverses direction exactly once it traversed the diffusion cloud's final diameter $d$ once to traverse it one more time.

\paragraph{Incorporating the slip-flow}
The mobile nucleus is moving with a Poiseuille profile, which we approximate to be oscillating in magnitude sinusoidally 
\begin{equation}
    u_{\text{n}}(R,t)= u_{\text{max}}\sin(\omega t)\left(1-\left[\frac{R}{R_{\text{in}}}\right]^2\right).
\end{equation}
Now, diffusing molecules emitted at a trapped nucleus at the inner tube boundary are much smaller than the nuclei and, therefore, co-move with the slip velocity profile.
\begin{equation}
    u_{m}(R_{\text{in}},t)= u_{\text{max}}\sin(\omega t)u_s.
\end{equation}
Thus, the relative velocity between the transversing nucleus and emitted molecules determines the velocity to transverse the diffusive cloud
\begin{equation}
    u(R,t)= u_{\text{max}}\sin(\omega t)\left(1-\left[\frac{R}{R_{\text{in}}}\right]^2\right)-u_{\text{max}}\sin(\omega t)u_s=u_{\text{max}}\sin(\omega t)\left(1-\left[\frac{R}{R_{\text{in}}}\right]^2-u_s\right).
\end{equation}
We focus on the flow velocity around the flow reversal point and approximate the time-dependent flow component further
\begin{equation}
    u_{\text{max}}\sin(\omega t)\approx u_{\text{max}}\omega t.
\end{equation}
And thus, the relative mobile nucleus trajectory  becomes,
\begin{equation}
   2\int_{0}^{\tau/2}u(R, s)ds=2\int_{0}^{\tau/2} u_{\text{max}}\omega s \left(1-\left[\frac{R}{R_{\text{in}}}\right]^2- u_s\right)ds = u_{\text{max}}\omega \frac{
   \tau^2}{4}\left(1-\left[\frac{R}{R_{\text{in}}}\right]^2- u_s\right).
\end{equation}
Plugging the result of the integral above back into Suppl.~Eq.~18 we find
\begin{figure}[t!]
	\centering
	\includegraphics[width=1\textwidth]{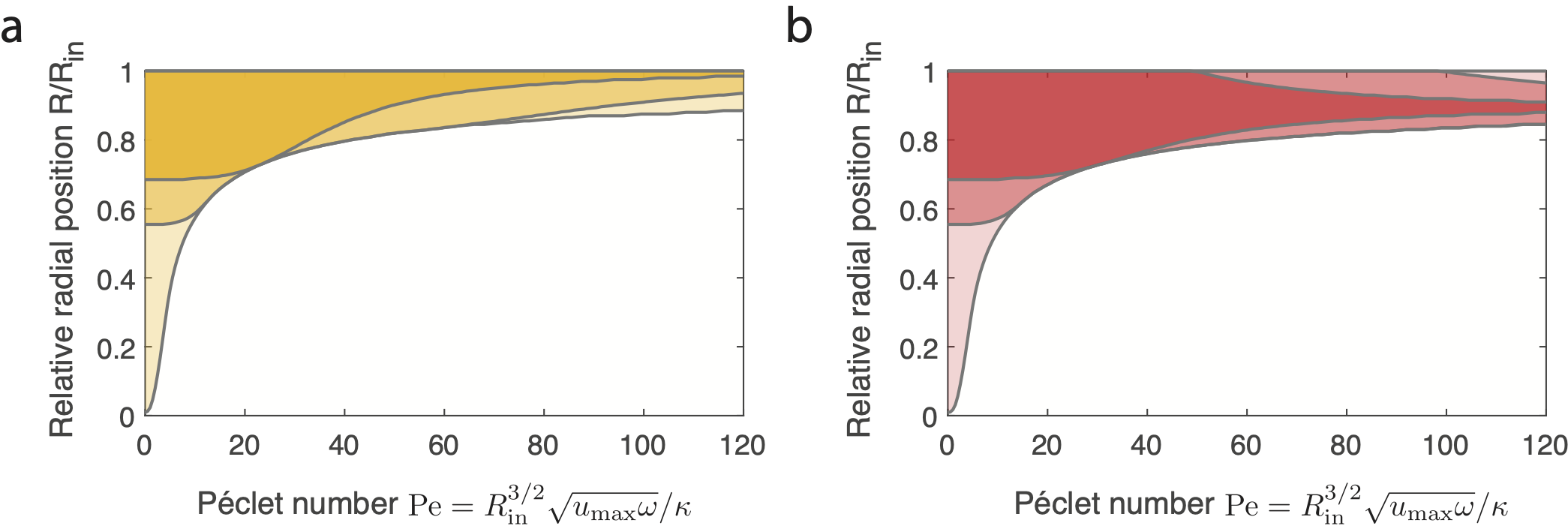}
	\caption{ Phase space of communication condition (a) without and (b) with $u_s=0.2$ slip velocity for $\tilde{\tau}\in[0.001,1]$, $\tilde{\tau}\in[0.04,0.2]$ and $\tilde{\tau}=0.1$ in increasing opacity.
}
	\label{fig:iso}
\end{figure}
\begin{equation}
    4R_{\text{in}}\sqrt{\frac{\kappa}{R_{\text{in}}^2} \tau-\left(1-\frac{R}{R_{\text{in}}}\right)^2}\geq u_{\text{max}}\omega \frac{\tau^2}{4}\left(1-\left[\frac{R}{R_{\text{in}}}\right]^2- u_s\right).
\end{equation}
Rearranging and squaring
\begin{equation}
 \left(\frac{16 R_{\text{in}}}{u_{\text{max}}\omega}\right)^2\left(\frac{\kappa}{R_{\text{in}}^2} \tau-\left(1-\frac{R}{R_{\text{in}}}\right)^2\right)\geq \tau^4 \left(1-\left[\frac{R}{R_{\text{in}}}\right]^2- u_s\right)^2.     
\end{equation}
Identifying $t_{\text{diff}}=\frac{R_{\text{in}}^2}{\kappa}$ as the time scale for diffusion across the tube and $t_{\text{adv}}=\sqrt{\frac{R_{\text{in}}}{u_{\text{max}} \omega}}$ as the time scale for flow driven nuclei transport, we can define the P\'eclet number as $\text{Pe}=t_{\text{diff}}/t_{\text{adv}}=\frac{R_{in}^{3/2}\sqrt{u_{max}\omega}}{\kappa}$. Now rescaling time in units of the time scale for diffusion by introducing $\tilde{\tau}=\tau/t_{\text{diff}}$ the inequality becomes a function of the dimensionless parameters $R/R_{\text{in}}$, $\tilde{\tau}$, $u_s$ and $\text{Pe}$ only,
\begin{equation}
\tilde{\tau}-\left(1-\frac{R}{R_{\text{in}}}\right)^2\geq \left(\frac{\text{Pe}\cdot\tilde{\tau}}{4}\right)^4\left[1-\left(\frac{R}{R_{\text{in}}}\right)^2-u_s\right]^2.
\end{equation}

\paragraph{Peristaltic flow}
Defining the contractions in a single peristaltic tube of length $L$  by
\begin{equation}
    a^2(z,t)= a_0^2+2a_0\alpha e^{i\left(\frac{2\pi z}{L}-\omega t\right)},
\end{equation}
such that we recover in the limit of small contraction amplitude $\alpha\to 0$ the classical peristaltic wave of the form
\begin{equation}
    a(z,t)= a_0(1+\alpha e^{i\left(\frac{2\pi z}{L}-\omega t\right)})+\mathcal{O}(\alpha^2),
\end{equation}
the cross-sectionally averaged flow velocity is given by 
\begin{eqnarray}
    \bar{u}(z,t) &=& \frac{Q(z_0,t)}{\pi a^2(z,t)}-\frac{1}{a^2(z,t)}\int_{z_0}^z dz'\frac{\partial a^2(z',t)}{\partial t}\nonumber\\
    &=&\frac{Q(z_0,t)}{\pi a^2(z,t)}+\frac{\omega}{a^2(z,t)}\frac{L}{2\pi}2a_0\alpha \left[e^{i\left(\frac{2\pi z}{L}-\omega t\right)}-e^{i\left(\frac{2\pi z_0}{L}-\omega t\right)}\right]\nonumber\\
    &=&\frac{Q(z_0,t)}{\pi a^2(z,t)}+\frac{\omega L \alpha e^{-i\omega t}}{\pi a_0+2\pi\alpha e^{i\left(\frac{2\pi z}{L}-\omega t\right)}} \left[e^{i\frac{2\pi z}{L}}-e^{i\frac{2\pi z_0}{L}}\right].
\end{eqnarray}
Taking the limit of small contraction amplitude $\alpha \ll1$, zero inflow at $z_0=0$, $Q(z_0,t)=0$ and long wavelength $z/L\ll1$ the expression for the flow velocity becomes
\begin{equation}
    \bar{u}(z,t)=\frac{\omega L \alpha e^{-i\omega t}}{\pi a_0} \frac{2\pi i z}{L}.
\end{equation}
So, indeed, the assumption of a sinusoidally varying flow velocity for the peristaltic flow velocity is satisfied.

\section*{SI Text 6: Scaling laws of signaling mechanisms}

In the main text, we compare the effectiveness of pigeon post with alternate signaling mechanisms, including (1)~pure advection, (2) pure diffusion and (3) diffusive relay. Here, we explicitly state how messaging time $t$ scales with distance covered $L$ for the four mechanisms. For pure advection, it is a linear motion given by \begin{equation}
t = \frac{L}{u_{max}}.
\label{pure_advection}
\end{equation}
\noindent For pure diffusion in one dimension \begin{equation}
t =  \frac{L^2}{2 \kappa},
\label{pure_diffusion}
\end{equation} given by Fickian diffusion. 
In diffusive relay such as in mitotic waves of \textit{Xenopus} \cite{Gelens.2014} or \textit{Drosophila melanogaster} \cite{Vergassola.2018} messages diffuse with diffusivity $\kappa$ between relay waypoints, with each waypoint itself being excitable doubling the messenger within $t_{\text{doubling}}$ giving way to a trigger wave propagating at constant speed \cite{Gelens.2014} 
\begin{equation}
t = \frac{L}{2}\sqrt{\frac{t_{\text{doubling}}}{\kappa}}.
\label{diffusion_relay}
\end{equation}
Note, that an enhancement of diffusivity by the effect of Taylor dispersion only is effective on length scales $L> \frac{R_{\text{in}}^2 u_{\text{max}}}{2\kappa}$, which for the physiological parameters considered here exceeds the entire length of the cell considered.

\section*{Supplementary Table 1: Physiological parameters}
\centering
\begin{tabular}{ |l | c | c | c |}
\hline
\multicolumn{4}{|c|}{Parameter List} \\
\hline
Parameter & Symbol & Value & Reference \\
\hline
Maximum flow velocity of nuclei on the central axis & $u_{max}(0)$ & $100-120\,\text{$\mu$m/s}$ & Experimental data \\
Diffusion constant of signaling molecule from the nuclei & $\kappa$ & $10\,\mu \text{m}^2\text{/s}$   & \cite{Gelens.2014} \\
Average inner tube radius & $R_{\text{in}}$ & $50\,\mu\text{m}$ & Experimental data, \cite{Kramar2021} \\
Average inter-nuclei distance & $n$    &$12\,\mu\text{m}$ & \cite{Gerber2022} \\
Doubling time of signaling molecule & $t_{\text{doubling}}$    &$10-100\,\text{s}$ & \cite{Gelens.2014}, \cite{Vergassola.2018} \\
Shuttle flow angular frequency & $\omega$    &$0.052\,\text{/s}$ & Experimental data,  \cite{Alim2013} ,\cite{Kramar2021} \\
Tube contraction period & $T$    &$120 \text{s}$ & Experimental data, \cite{Alim2013},\cite{Kramar2021}\\
\hline
\end{tabular}